\documentstyle[12pt,twoside]{article}
{\catcode `\@=11 \global\let\AddToReset=\@addtoreset}
\AddToReset{equation}{section}
\renewcommand{\theequation}{\thesection.\arabic{equation}}  

\def\greaterthansquiggle{\raise.3ex\hbox{$>$\kern-.75em\lower1ex\hbox{$\sim$}}}
\def\lessthansquiggle{\raise.3ex\hbox{$<$\kern-.75em\lower1ex\hbox{$\sim$}}}
\newcommand{\beq}{\begin{equation}}
\newcommand{\eeq}{\end{equation}}
\newcommand{\beqa}{\begin{eqnarray}}
\newcommand{\eeqa}{\end{eqnarray}}
\newcommand{\beqan}{\begin{eqnarray*}}
\newcommand{\eeqan}{\end{eqnarray*}}
\newcommand{\ba}{\begin{array}}
\newcommand{\ea}{\end{array}}
\newcommand{\no}{\nonumber}

\newcommand{\ra}{\rightarrow}
\newcommand{\Ra}{\Rightarrow}

\newcommand{\vp}{\varphi}

\newcommand{\F}{{\cal F}}

\newcommand{\cL}{{\cal L}}

\newcommand{\cS}{{\cal S}}

\newcommand{\dfrac}{\displaystyle \frac}

\def\QED{\\ {\hspace*{\fill}{\vrule height 1.8ex width 1.8ex }\quad}
    \vskip 0pt plus20pt}

\def\nz{\ifmmode {I\hskip -3pt N} \else {\hbox {$I\hskip -3pt N$}}\fi}
\def\zz{\ifmmode {Z\hskip -4.8pt Z} \else
       {\hbox {$Z\hskip -4.8pt Z$}}\fi}
\def\qz{\ifmmode {Q\hskip -5.0pt\vrule height6.0pt depth 0pt
       \hskip 6pt} \else {\hbox
       {$Q\hskip -5.0pt\vrule height6.0pt depth 0pt\hskip 6pt$}}\fi}
\def\rz{\ifmmode {I\hskip -3pt R} \else {\hbox {$I\hskip -3pt R$}}\fi}
\def\cz{\ifmmode {C\hskip -4.8pt\vrule height5.8pt\hskip 6.3pt} \else
       {\hbox {$C\hskip -4.8pt\vrule height5.8pt\hskip 6.3pt$}}\fi}

\def\au{{\setbox0=\hbox{\lower1.36775ex%
\hbox{''}\kern-.05em}\dp0=.36775ex\hskip0pt\box0}}
\def\ao{{}\kern-.10em\hbox{``}}

\normalsize
\begin{document}
\bibliographystyle{plain}
\begin{titlepage}
\begin{flushright}
UWThPh-1997-16 \\
\today
\end{flushright}
\vspace{1cm}
\begin{center}
{\Large \bf
Late time behaviour of the maximal slicing of the Schwarzschild black 
hole}\\[1 cm]
R. Beig  \\
Institut f\"ur Theoretische Physik \\
Universit\"at Wien \\
Boltzmanngasse 5, A--1090 Wien, Austria\\
Fax: ++43-1-317-22-20, E-mail: BEIG@PAP.UNIVIE.AC.AT \\[4pt]
and \\[4pt]
N. \'O Murchadha* \\
Physics Department \\
University College Cork \\
Cork, Ireland\\
Fax: +353-21-276949, E-mail: NIALL@BUREAU.UCC.IE\\
and\\
Erwin Schr\"odinger International Institute\\
for Mathematical Physics\\
Boltzmanngasse 9, A-1090 Vienna, Austria
\vfill
{\bf Abstract} \\
\end{center}

A time-symmetric Cauchy slice of the extended Schwarzschild spacetime
can be evolved into a foliation of the $r > 3m/2$-region of the
spacetime by maximal surfaces with the requirement that time runs
equally fast at both spatial ends of the manifold. This paper studies
the behaviour of these slices in the limit as proper time-at-infinity
becomes arbitrarily large and gives an analytic expression for the
collapse of the lapse.
\vfill
\noindent 
*) Partially supported by Forbairt Grant SC/96/225.
\end{titlepage}

\section{Introduction}
In this work we study the time function $\tau$ on the Schwarzschild
black hole spacetime having the following properties:
\begin{enumerate}
\item[(i)] The level sets of $\tau$ result from evolution of a time-
symmetric Cauchy
slice of Schwarzschild by maximal surfaces under the additional requirement
that proper time for asymptotic observers at infinity, which are
at rest relative to the slicing, runs equally fast at both spatial ends.
\item[(ii)] The time function $\tau$ is zero on the time-symmetric slice
and coincides with the proper time of the infinite observers.
(This means that
$\alpha$, the lapse of the time function goes to one at both infinities
along each slice.)
\end{enumerate}
Note that (i) is really a property only of the slicing defined
by $\tau$ rather than $\tau$ itself. This time function which has first 
been considered in \cite{Es,Re} has two key properties: The first property
is that $\tau$ takes all real values or, in other words, the future
singularity at $r = 0$ does not prevent $\tau$ from assuming arbitrarily
large positive values (and similarly for the past).  It is believed
that this property holds on vacuum spacetimes more general than Schwarzschild.
Here it is important to realize that such spacetimes are not ``given''
to us: rather they have to be generated by a Cauchy problem: One first
constructs regular asymptotically flat initial data, satisfying the
vacuum constraints, say maximal, and then tries to evolve these in
time by analytical or numerical means. Doing this involves an a priori
choice of gauge which in particular implies that the resultant globally
hyperbolic spacetime comes already equipped with a specific time
function. Suppose the initial data has a future-trapped surface. Then, 
by the Penrose singularity theorem \cite{Pe1}, any Cauchy evolved
spacetime is singular in the sense of having future-incomplete 
null geodesics. (Similar conclusions, but in both the future and the past
direction, hold when the initial data has an outer-trapped surface 
\cite{To} or when the topology is nontrivial e.g. in the sense that
there is more than one asymptotic end \cite{Ga}.) Many maximal initial
data sets having one of these properties exist (for trapped surfaces, 
see \cite{Be}). Now there is the conjecture, due to Moncrief and Eardley
\cite{Mo}, that if one evolves the initial data in a gauge where the
whole slicing is maximal and $\tau$ is proper time at infinity, the
evolution should be extendable to arbitrarily large values of $\tau$,
irrespective of whether singularities form or not.
This global existence result, if true, would, in spirit at least,
go a long way toward settling in the affirmative the Penrose Cosmic
Censorshop hypotheses \cite{Pe2} in the case of asymptotically flat
data. The spacetime evolving in the way described, in the Schwarzschild
case, has the second property that it is in fact extendable: there are
no maximal Cauchy slices of Schwarzschild reaching radii less than or
equal to $r = 3m/2$. Thus maximal slices of Schwarzschild ``avoid
the singularity at $r = 0$''. It is this last property which numerical
relativists expect to be true for evolutions of more general initial
data and which is clearly desirable if numerical codes based on
maximal slicings are used. 

Take any observer at rest relative to the 
slicing defined by $\tau$ (``Eulerian observer''). Then 
$\int \alpha d\tau$ along the trajectory of that observer is her or 
his proper time.
Since proper time is finite as the slicing approaches the limiting
maximal slice at $r = 3m/2$, we must have $\int \alpha d\tau < \infty$,
and thus $\lim_{\tau \ra \infty} \alpha(\tau) = 0$ (``collapse of the
lapse'' \cite{Yo1}). Our main result is that, along the Eulerian observers
going through the bifurcation 2-sphere,
\beq
\alpha(\tau) \sim \frac{4}{3 \sqrt{2}} \exp \frac{4A}{3 \sqrt{6}}
\exp \left( - \frac{4\tau}{3 \sqrt{6} \; m}\right) \mbox{ as }
\tau \ra \infty,
\eeq
where the constant $A$ is given by Equ. (3.41). The
exponent in (1.1) has been estimated before \cite{Es,Yo2} by a mixture 
of numerical and model calculations. The estimate in \cite{Yo2} of this
exponent is 1.82 which agrees quite closely with our exact $\frac{3
\sqrt{6}}{4} \sim 1.83$.
We hope that this result will be
useful for the numerists as an accurate test for codes based on maximal
slicings. An extension of the work here to the late time behaviour of
$\alpha$ along the trajectories of arbitrary Eulerian observers
will appear
elsewhere \cite{Ca}.

Our plan is as follows: In \S~2 we review some generalities on lapse 
functions and foliations. Then we give a precise definition of the
time function under study. In \S~3 we perform the asymptotic analysis
leading to Equ. (1.1). In Appendix A we essentially rederive the
Schwarzschild metric in terms of spherically symmetric, maximal Cauchy
data. In Appendix B we prove a calculus lemma which is basic for our
analysis.

\section{Generalities}
Let $(M,ds^2)$ be a globally hyperbolic spacetime and $\tau : M \ra
{\bf R}$ a time function, i.e. a function the level sets of which form
a foliation $\F_\tau$ of $M$ by Cauchy surfaces $\cong \Sigma$. Then
the function $\alpha : M \ra {\bf R}$ defined by
\beq
\alpha := (-(\nabla \tau)^2)^{-1/2}
\eeq
is called the {\bf lapse} of $\F_\tau$. The reason for this name is that
$\alpha$ measures ``lapse of proper time'' along trajectories normal to
the leaves of $\F_\tau$ as a function of $\tau$. To make this explicit, 
define the vector field $\tau^\mu$ by
\beq
\tau^\mu = - \alpha^2 \nabla^\mu \tau \Ra \tau^\mu \nabla_\mu \tau = 1
\eeq
which is timelike and future (i.e. increasing $\tau$)
pointing. The vector $\tau^\mu$ yields an
orthogonal decomposition of $M$ as $M = {\bf R} \times \Sigma$, as
follows: Construct a diffeomorphism $\vp : {\bf R} \times \Sigma$, i.e.
$\vp : (\lambda,y^i) \in {\bf R} \times \Sigma \mapsto x^\mu =
\vp^\mu_\lambda(y^i) \in M$, by
\beqa
\dot \vp^\mu_\lambda(y) = \frac{d}{d\lambda} \vp^\mu_\lambda(y) &=&
\tau^\mu(\vp_\lambda(y)) \no \\
\vp_0 &=& \tau^{-1}(0).
\eeqa
Thus $\lambda$, viewed as a function on $M$, coincides with $\tau$. 
We will, by abuse of notation, use the same letter $\tau$ for $\lambda$
viewed in this way. Furthermore, since $\tau^\mu$ is normal to the
leaves of $\F_\tau$, the lines of constant $y^i$ are orthogonal 
trajectories, in other words
\beq
\dot \vp^\mu_\tau(y) \vp^\nu_{\tau,i}(y) g_{\mu\nu}(\vp_\tau(y)) = 0.
\eeq
Consequently, in the $(\tau,y^i)$-coordinates, the metric takes the form
\beqa
\vp^*_\tau(ds^2) &=& \dot \vp^\mu_\tau \dot \vp^\nu_\tau g_{\mu\nu} d\tau^2
+ \vp^\mu_{\tau,i} \vp^\nu_{\tau,j} g_{\mu\nu} dy^i dy^j \no \\
&=& g_{\tau \tau}(\tau,y) d\tau^2 + g_{ij}(\tau,y) dy^i dy^j ,
\eeqa
where $g_{ij}$ is the induced metric on the leaves and
\beq
g_{\tau\tau}(\tau,y) = - \alpha^2(\vp_\tau(y)).
\eeq
Thus, along $y^i =$ constant, the proper time $s$ is given by
\beq
s = \int \alpha(\vp_{\tau'}(y)) d\tau'.
\eeq
Note that, when $\tau'$ is another time function giving the same
foliation, i.e. $\tau' = \tau'(\tau)$, the lapse $\alpha$ changes
according to $\alpha' = \left(\dfrac{d\tau'}{d\tau}\right)^{-1}\alpha$.
Suppose now we are given another vector field $\xi^\mu$ on $M$.
This can be uniquely decomposed
\beq
\xi^\mu = N n^\mu + X^\mu, \qquad X^\mu n_\mu = 0
\eeq
where $n^\mu = - \alpha \nabla^\mu \tau$, is the future normal of 
$\F_\tau$. To distinguish $N$ from $\alpha$, we call $N$ the 
{\bf boost function} of $\xi^\mu$ relative to $\F_\tau$. If $N$ is non-zero
on some leaf $\Sigma_{\tau_0}$, it can be viewed as the restriction to
$\Sigma_{\tau_0}$ of the lapse of the time function $t'$ obtained by
$\xi^\mu \nabla_\mu t' = 0$, $t'|_{\Sigma_{\tau_0}} =$ const.

We have the relation
\beq
N = \alpha \; \xi^\mu \nabla_\mu \tau ,
\eeq
which is of course trivial in the present context, but will be extremely
useful in our computation of the lapse $\alpha$ of a maximal foliation
of the extended Schwarzschild spacetime, where $\xi^\mu$ can be chosen as
the "static" Killing vector.

We now recall some features of Schwarzschild which are used in our
construction. In the exterior region $r > 2m > 0$ we have
\beq
ds^2 = - \left( 1 - \frac{2m}{r}\right)dt^2 + \left( 1 - 
\frac{2m}{r}\right)^{-1} dr^2 + r^2 d\Omega^2, \qquad
- \infty < t < \infty.
\eeq
$ds^2$ can be smoothly extended across $r = 2m$ to the Kruskal spacetime
$M$ on which $r$ is a globally defined function $r : M \ra {\bf R}^+$,
which has saddle points at $\cS$, the bifurcation 2-sphere of the horizon.
The Killing vector field $\partial/\partial t$ extends to a global
Killing vector field $\xi^\mu$ on $M$ which is spacelike in the interior,
i.e. black and white hole, regions, null on the horizon and zero on
$\cS$. 
Both the black hole region and the right exterior region can be written in 
the form (2.10) with the understanding that the functions $(\theta,\vp)$
and $r$ together with the retarded Eddington--Finkelstein coordinate
\beq
u = t - r - 2m \lg |r - 2m|
\eeq
covers both regions and the horizon at $r = 2m$.
 The function $t$ goes to $\infty$ at the right 
component (where ``right'' refers to the original unextended spacetime)
and goes to $- \infty$ at the left horizon. The set where $t$ vanishes
is the union of $S$, the original $t = 0$-spacelike hypersurface (extended
in the obvious way to the left exterior region) and the timelike, totally
geodesic cylinder $\Gamma$, which is ruled by timelike radial geodesics
through $\cS$ which are orthogonal to $S$, and
which hit the singularity as $r \ra 0$. Since $r$ is constant along the 
trajectories of $\xi^\mu$ and $r$ is, by (2.10), an ``areal radius'',
it follows that every spherically symmetric spacelike slice has a spherical
minimal surface (a ``throat'') exactly where it is tangential to $\xi^\mu$
(which of course can only happen in the interior and it necessarily has
to happen there for slices leaving to the other (left) exterior region).

Consider the function $h(r,C)$ given by
\beq
h(r,C) = - \int_{r_C}^r \frac{C}{(1 - 2m/x)(x^4 - 2mx^3 + C^2)^{1/2}} dx,
\eeq
where the integral is to be understood in the Cauchy-principal-value sense
for $r > 2m$
and where $0 < C < 3 \frac{\sqrt{3}}{4}\; m^2$, $r > r_C$ and $r_C$ is the
unique root of $P(x) = x^4 - 2mx^3 + C^2$ for this range of $C$ in the
interval $3m/2 < r_C < 2m$. For $x > r_C$, we have $P(x) > 0$. Thus
$h(r,C) + r + 2m \lg |r-2m|$ depends smoothly on $(r,r_C)$. We easily
infer that
\beq
t = h(r,C)
\eeq
defines, for each fixed $C$, a spacelike slice $\Sigma_C$ which smoothly
extends to the black hole, where it intersects $\Gamma$ at $r = r_C$.

In order to see that this surface extends smoothly and symmetrically
through $\Gamma$, we use for $r < 2m$ the parameter
\beq
\ell(r) = \int_{r_C}^r \frac{x^2 dx}{[P(x)]^{1/2}},
\eeq
which is the proper distance along the slice, as can either be seen from
(2.10,12,13) or from Appendix A. Then, from (2.12,14) we have the system
of ODE's
\beqa
\frac{dh}{d\ell} &=& - \frac{C}{r^2 - 2mr} \no \\
\frac{d^2r}{d\ell^2} &=& \frac{m}{r^2} - \frac{2C^2}{r^5} ,
\eeqa
with $h(0) = 0$, $r(0) = r_C$, $\frac{dr}{d\ell}(0) = 0$, which is
regular at $\ell = 0$.  Thus the function $r$ along the slice
is symmetric with respect to $\ell = 0$ and smooth. This implies that
$dr/d\ell = (1 - 2m/r + C^2/r^4)^{1/2}$ is antisymmetric.

Next we observe that the level sets of $\sigma = t - h(r,C)$, for fixed
$C$ in the allowed range, give rise to maximal surfaces on the Kruskal
manifold, i.e. they satisfy
\beq
\nabla^\mu([- (\nabla \sigma)^2]^{-1/2} \nabla_\mu \sigma) = 0.
\eeq
The function $\sigma$ is not the time function of interest to us (in fact:
$\sigma$ being not differentiable at $r_C$ it does not define a global
foliation). Rather this local foliation arises from moving a given
maximal slice, say $\sigma = 0$, along the flow of
$\xi^\mu = (\partial/\partial t)^\mu$. 
The function $N = [-(\nabla \sigma)^2]^{-1/2}$ is nothing but 
the boost function of $\partial/\partial t$ relative to 
$\sigma = 0$. In fact, in the explicit solution of (2.14) due to
Reinhart \cite{Re}, he first, essentially by guessing, finds $N$ to be
\beq
N = \left( 1 - \frac{2m}{r} + \frac{C^2}{r^4}\right)^{1/2},
\eeq
and from this (2.12) can be inferred. 
For a more illustrative derivation from the initial-value
point of view see Appendix A. Note that $N$ as a function of $\ell$ is
antisymmetric relative to $\ell = 0$.

We now claim that the surface $t = h(r,C')$ lies everywhere in the future of
$t = h(r,C)$ when $C' > C;$ and that $t = h(r,C)$ lies to the future of
$S = \Sigma_0$. Amusingly we are unable to see this from
the explicit integral (2.12). Instead, one first computes
$\frac{d}{dC} r_C$ from
\beq
r_C^4 - 2m r_C^3 + C^2 = 0 ,
\eeq
to yield
\beq
\frac{dr_C}{dC} = - \frac{2C}{4r_C^3 (1 - 3m/2r_C)} < 0.
\eeq
Thus the claimed behaviour is true at least along the throat. Next
observe that our slices are asymptotically flat at both spatial ends
and that $t_\infty(C) = \lim_{r \ra \infty} h(r,C)$ exists. Suppose
that $h(R,C) = h(R,C')$ for some $R > r_C$ to the right of $\Gamma$.
Then, by the symmetry w.r.t. $\Gamma$ this would have to happen also
to the left of $\Gamma$. Thus we would have a lens-shaped region
spanned by two maximal slices.
But this, by an elegant argument due to
Brill and Flaherty \cite{Br}, is impossible, except if the two slices are
identical, which they are not in our case. This
argument continues to be valid for $R = \infty$. Thus $t(r,C)$
monotonically increases with $C$ for fixed $r$, and so does
$t_\infty(C)$.

It follows that the equation $t = h(r,C)$ can be solved for $C$
to yield a smooth time function defined on the $r < 3m/2$-subset of the
part of Kruskal lying in the future of the Cauchy slice $S$.
$C$ gives the foliation we are interested in, but it is not yet the
time function we want: rather this is obtained from eliminating $C$
in terms of $\tau$ in the relation
\beq
\tau = t_\infty(C).
\eeq

Suppose we had started with the Cauchy slice $t = 0$ which, being
time-symmetric, is in particular maximal and evolve it into a maximal
slicing by a lapse function $\alpha$ going to 1 at both spatial ends.
This is possible in a unique way (see \cite{Ba}).
Then the resultant time function is spherically symmetric and
symmetric w.r.t. $\Gamma$, and so it has to coincide with the one
obtained above. In particular it follows that our $\tau$ can be smoothly
extended to negative values of $\tau$ which would have been very
non-obvious from the explicit formula (2.12).

We next compute the lapse function $\alpha$ of $\tau$. Using (2.9) this
involves computing
\beq
(\xi^\mu \nabla_\mu \tau)^{-1} = \frac{dC}{d\tau} 
\left. \frac{\partial h}{\partial C} \right|_r.
\eeq
Note that (2.21) blows up at $r = r_C$ but in such a way that
\beq
\alpha = (\xi^\mu \nabla_\mu \tau)^{-1} N
\eeq
has a smooth limit as $r \ra r_C$, as it has to be. Using formula (B.12),
there results
\beq
\alpha = \left(\frac{d\tau}{dC} \right)^{-1} \frac{1}{2} \left[
\frac{1}{r - 3m/2} - \int_{r_C}^r \frac{x(x-3m)dx}
{(x - 3m/2)^2 [x^4 - 2mx^3 + C^2]^{1/2}}\right],
\eeq
with
\beq
\frac{d\tau}{dC} = \int_{r_C}^\infty \frac{x(x-3m)dx} {(x - 3m/2)^2 
[x^4 - 2mx^3 + C^2]^{1/2}}.
\eeq
Note that $N$ and $\alpha$ are linearly independent radial solutions of
\beq
(\Delta - K_{ij} K^{ij})f = 0,
\eeq
where $N$ goes to 1 at the right infinity and to $-1$ at the left one
whereas $\alpha$ goes to 1 at both ends. 

We are interested in studying $\alpha$ along the trajectories of
Eulerian observers. This
requires choosing a coordinate $\rho = \rho(r,\tau)$ the level surfaces
of which are timelike cylinders orthogonal to our slicing.
(One such timelike cylinder is 
already known, namely $\Gamma$ given by $r = r_C$.) Such a coordinate
can be found without any calculation. Recall that maximal slicings 
preserve spatial volumes along Eulerian observers. Thus a suitable 
coordinate will be the ``volume radius'' on each slice, defined by
\beq
\rho^3(r,\tau) = 3 \int_{r_{C(\tau)}}^r \frac{x^4
dx}{[x^4-2mx^3+C^2]^{1/2}},
\eeq
using that
the spatial metric on each slice has the form (see Appendix A)
\beq
g_{ij} dx^i dx^j = \left(1 - \frac{2m}{r} + \frac{C^2}{r^4} \right)^{-1}
dr^2 + r^2 d\Omega^2.
\eeq
In the coordinates $(t,\rho,\theta,\vp)$, the Schwarzschild metric
for $r > 3m/2$ reads
\beq
ds^2 = - \alpha^2 d\tau^2 + 
\left(\frac{\rho}{r}\right)^4 d\rho^2 + r^2 d\Omega^2
\eeq
where $r = r(\rho,\tau)$ is given implicitly by (2.26) and $\alpha$ by
(2.23). (To check (2.28) explicitly one should first observe that
$C \left.\frac{\partial h}{\partial C}\right|_r = \rho^2 \left.\frac{\partial 
\rho}{\partial C}\right|_r$.)

Note that, as $C$ approaches $\sqrt{27/16} \; m^2$,  $r_C$ approaches
the value $3m/2$, since
\beq
P(x) = x^4 - 2mx^3 + C^2 = \left(x - \frac{3m}{2} \right)^2
\left(x^2 + mx + \frac{3m^2}{4} \right) +
O\left( \left(x - \frac{3m}{2} \right)^2\right).
\eeq
Equ. (2.29) also shows that $r_C$ approaches a double root of $P(x)$ as
$C \ra \sqrt{27/16} \; m^2$. Thus, as one lets $\tau$ tend to infinity for
fixed $\rho$, the function $r$ approaches $3m/2$. In that sense the
slices approach the limiting maximal slice at $r = 3m/2$. We are
interested in estimating $\alpha$ in that limit. For simplicity we will
confine ourselves to $\rho = 0$, i.e. the throat $\Gamma$.

\section{The late time analysis}
It is convenient to replace the parameter $C$ by $\delta$ defined by
\beq
\delta = r_C - \frac{3m}{2}, \qquad
r_C^4 - 2mr_C^3 + C^2 = 0.
\eeq
As $C$ ranges between 0 and $3\; \frac{{\sqrt 3}}{4} \; m^2$, $\delta$
ranges monotonically from $m/2$ to 0. Using the rescaled quantities
\beq
\bar C = \frac{C}{m^2}, \qquad
\bar \tau = \frac{\tau}{m}, \qquad
\bar \delta = \frac{\delta}{m},
\eeq
we find that
\beq
\bar \tau(\bar \delta) = - \bar C \int_{3/2 + \bar\delta}^\infty
\frac{y dy}{(y-2)(y^4-2y^3+\bar C^2)^{1/2}},
\eeq
where
\beq
\bar C = \left(\bar \delta + \frac{3}{2} \right)^{3/2}
\left(\frac{1}{2} - \bar \delta \right)^{1/2}.
\eeq
We have the
\paragraph{Lemma:}
\beqa
\bar \tau(\bar \delta) &=& - \frac{3\sqrt{6}}{4} \ln \bar \delta +
\frac{3 \sqrt{6}}{4} \ln |18(3 \sqrt{2} - 4)| -
2 \ln \left|\frac{3 \sqrt{3} - 5}{9 \sqrt{6} - 22} \right| + 
O(\bar \delta) \no \\
&=& - \frac{3 \sqrt{6}}{4} \ln \bar \delta + A + O(\bar \delta) \qquad
\mbox{as } \bar \delta \ra 0.
\eeqa
\paragraph{Proof:} First note that
\beq
\frac{d}{d \bar C} \left[\frac{\bar C}{(y^4 - 2y^3 + \bar C^2)^{1/2}} \right]
= \frac{y^3(y-2)}{(y^4 - 2y^3 + \bar C^2)^{3/2}}.
\eeq
Thus, from the mean value theorem,
\beq
\left| \frac{y}{y-2} \left(\frac{\bar C}{(y^4-2y^3+\bar C^2)^{1/2}}
- \frac{\sqrt{27/16}}{(y^4-2y^3+ 27/16)^{1/2}} \right)\right| \leq
\frac{\sqrt{27} \; \bar \delta^2 y^4}{(y^4-2y^3+\bar C^2)^{3/2}}
\eeq
where we have used
\beq
\sqrt{27/16} - \bar C \leq \sqrt{27} \; \bar \delta^2.
\eeq
(3.7) is valid 
for $y \neq 2$ but, by continuity, also for $y = 2$. We will find it
convenient to sometimes express $\bar C$ in terms of $\bar \delta$,
using (3.4). Writing
\beq 
Q(s) = s^2\left(s^2 + 4s + \frac{9}{2}\right) - \bar \delta^2
\left(\bar \delta^2 + 4 \bar \delta + \frac{9}{2}\right).
\eeq
Equ. (3.3) can, after substituting $s = y - 3/2$, be written as
\beq
\bar \tau = \left(\bar \delta + \frac{3}{2}\right)^{3/2}
\left(\frac{1}{2} - \bar \delta\right)^{1/2} \int_{\bar \delta}^\infty
\frac{(s + 3/2)ds}{(1/2 - s)[Q(s)]^{1/2}}.
\eeq
It is elementary to see that, for $s \geq \bar \delta$,
\beq
0 \leq \frac{9}{2} (s^2 - \bar \delta^2) \leq Q(s) \leq (s^2 - 
\bar \delta^2)\left[\frac{9}{2} + 2s(4+s) \right]
\eeq
which, using $\sqrt{1+x} \leq 1 + x/2$ for $x \geq 0$, implies
\beq
\left|\frac{1}{[Q(s)]^{1/2}} - 
\frac{1}{[\frac{9}{2}(s^2-\bar \delta^2)]^{1/2}} \right| \leq
\frac{\frac{2s}{9}(4+s)}{[\frac{9}{2}(s^2 - \bar \delta^2)]^{1/2}}.
\eeq
The estimate (3.7) now takes the form
$$
\left|\frac{s + 3/2}{1/2 - s} \left( 
\frac{(\bar \delta + 3/2)^{3/2}(1/2 - \bar \delta)^{1/2}}{[Q(s)]^{1/2}}
- \frac{\sqrt{27/16}}{[s^2(s^2+4s+9/2)]^{1/2}}
\right) \right| \leq 
\frac{\sqrt{27} \; \bar \delta^2(s + 3/2)^4}{[Q(s)]^{3/2}} 
$$
\beq
\leq \frac{\sqrt{27}\; \bar \delta^2(s + 3/2)^4}{[\frac{9}{2}(s^2 -
\bar \delta^2)]^{3/2}}.
\eeq
The inequalities (3.12,13) are the basic estimates we will be using. We
now split the integration domain in (3.10) as follows:
\beq
\bar \delta \leq s \leq \sqrt{\bar \delta/2 }, \qquad
\sqrt{\bar \delta/2} \leq s \leq \infty
\eeq
and write
\beq
\bar \tau = \bar \tau_1 + \bar \tau_2 
\eeq
accordingly. We furthermore define ($0 < \bar \delta < 1/2$)
\beqa
\bar \tau^0_1 &=& \sqrt{27/16} \int_{\bar \delta}^{\sqrt{\bar\delta/2}}
\frac{s + 3/2}{(1/2 - s)[\frac{9}{2}(s^2-\bar\delta^2)]^{1/2}}ds \\
\bar \tau^0_2 &=& \sqrt{27/16} \int_{\sqrt{\bar\delta/2}}^\infty
\frac{s + 3/2}{(1/2 - s)[s^2(s^2 + 4s + 9/2)]^{1/2}} ds .
\eeqa
Equ. (3.17) is in the principal-value sense at $s = 1/2$. These integrals
can be explicitly computed using the formulas (see e.g. \cite{Te})
\beqa
\int \frac{dx}{x^2\sqrt{x^2 - \bar \delta^2}} &=&
\frac{\sqrt{x^2 - \bar \delta^2}}{x \bar \delta^2}, \qquad
x > \bar \delta > 0, \\
\int \frac{dx}{x^2\sqrt{x^2 - \bar{\delta}^2}} &=&
\ln  |x + \sqrt{x^2 - \bar \delta^2}|, \qquad
x > \bar \delta > 0,
\eeqa
\beq
\int \frac{dx}{x \sqrt{ax^2 + bx + c}} = \frac{1}{\sqrt{c}} \ln
\frac{|-2 \sqrt{c(ax^2+bx+c)} + 2c +bx|}{2x}, \qquad c > 0.
\eeq
Using $\dfrac{s+3/2}{s(1/2 - s)} = \dfrac{3}{s} - \dfrac{4}{s - 1/2}$,
there results after straightforward manipulations
\beqa
\bar \tau^0_1 &=& - \frac{3 \sqrt{6}}{4} \ln \bar \delta + 
\frac{3 \sqrt{6}}{4} \ln \sqrt{\bar \delta/2} + o(1) \mbox{ as }
\bar \delta \ra 0 \\
\bar \tau^0_2 &=& - \frac{3 \sqrt{6}}{4} \ln \sqrt{\bar \delta/2} + 
\frac{3 \sqrt{6}}{4} \ln 2 + \frac{3 \sqrt{6}}{4} \ln \left|
\frac{18}{4 + 3\sqrt{2}}\right| - 2 \ln \left|
\frac{3 \sqrt{3}-5}{9\sqrt{6} - 22}\right| + o(1) \no \\
&& \mbox{ as } \bar \delta \ra 0 .
\eeqa
Next we have to estimate the remainders. We have
\beq
\Delta \bar \tau_1 = \int_{\bar\delta}^{\sqrt{\bar\delta/2}}
\left[ \frac{\bar C(\bar\delta)}{[Q(s)]^{1/2}} -
\frac{\sqrt{27/16}}{[\frac{9}{2}(s^2 - \bar\delta^2)]^{1/2}}\right]ds.
\eeq
Using $\bar C(\bar\delta) = \sqrt{27/16} + O(\bar\delta^2)$ and (3.11,12),
this has 
\beq
|\Delta \bar \tau_1| \leq \mbox{const}
\int_{\bar\delta}^{\sqrt{\bar\delta/2}} 
\frac{s ds}{\sqrt{s^2 - \bar\delta^2}} = O(\bar \delta^{1/2}).
\eeq
Next
\beq
\Delta \bar \tau_2 = \int_{\sqrt{\bar\delta}/2}^\infty
\frac{s + 3/2}{1/2 - s}
\left[ \frac{\bar C(\bar\delta)}{[Q(s)]^{1/2}} -
\frac{\sqrt{27/16}}{[s^2(s^2 + 4s + 9/2)]^{1/2}}\right] ds.
\eeq
By (3.13), this has a bound of the form
\beq
|\Delta \bar\tau_2| \leq \mbox{const} \;  \bar\delta^2
\int_{\sqrt{\bar\delta/2}}^\infty \frac{(s + 3/2)^4}
{[s^2(s^2 + 4s + 9/2)]^{3/2}} ds = \mbox{const} \; \bar\delta^2 I.
\eeq
The integral $I$ in (3.26) can be further split as $I = I_2 + I'_2$, where
\beqa
I_2 &=& \int_{\sqrt{\bar\delta/2}}^1 \frac{(s + 3/2)^4}
{[s^2(s^2 + 4s + 9/2)]^{3/2}} ds \leq \mbox{const} 
\int_{\sqrt{\bar\delta/2}}^1 
\frac{ds}{(s^2 - \bar\delta^2)^{3/2}} \no \\
&=& \mbox{const} \frac{1}{\bar\delta^2}
\int_{\sqrt{2/\bar\delta}}^{1/\bar\delta} \frac{df}{(f^2 - 1)^{3/2}} =
O\left(\frac{1}{\bar\delta}\right).
\eeqa
Now
\beq
I'_2 = \int_1^\infty \frac{(s + 3/2)^4}{[s^2(s^2+4s+9/2)]^{3/2}} ds \leq
\mbox{const} \int_1^\infty \frac{ds}{s^2} < \infty.
\eeq
Thus $\Delta \bar \tau_2 = O(\bar\delta)$. Putting all this together implies
\beq
\bar\tau(\bar \delta) = - \frac{3 \sqrt{6}}{4} \ln \bar\delta + A + o(1)
\mbox{ as } \bar\delta \ra 0,
\eeq
which is not quite good enough. From Equ. (B.12) in the limit that $r$
goes 
to infinity and
\beq
2 \bar C \frac{d \bar C}{d \bar\delta} = - 4 \bar\delta\left(\bar\delta +
\frac{3}{2}\right)^2,
\eeq
we see that
\beqa
\frac{d\bar\tau}{d\bar\delta} &=& \frac{\bar\delta(\bar\delta + 3/2)^{1/2}}
{(1/2 - \bar\delta)^{1/2}} \int_{\bar\delta}^\infty
\frac{(s+3/2)(s-3/2)}{s^2[Q(s)]^{1/2}} ds \no \\
&=& \sqrt{3}\; \bar\delta \int_{\bar\delta}^1 
\frac{s^2 - 9/4}{s^2} \frac{ds}{[Q(s)]^{1/2}} + O(\bar\delta) =
\sqrt{3} \; \bar \delta \; J + O(\bar\delta).
\eeqa
$J$ can in turn be split as $J = J^0 + \Delta J$, where
\beqa
J^0 &=& \int_{\bar\delta}^1 \frac{s^2 - 9/4}{s^2} \;
\frac{ds}{[\frac{9}{2} (s^2 - \bar \delta^2)]^{1/2}} = \no \\
&=& \sqrt{2/9} \int_1^{1/\bar\delta} \frac{df}{\sqrt{f^2-1}} -
9/4\; \sqrt{2/9} \; \frac{1}{\bar\delta^2} \int_1^\infty
\frac{df}{f^2 \sqrt{f^2 - 1}} + O\left(\frac{1}{\bar\delta}\right) \no \\
&=& O(\ln \bar\delta) - \sqrt{9/8} \; \frac{1}{\bar\delta^2} \cdot 1.
\eeqa
Finally,
\beq
\Delta J = \int_{\bar\delta}^1 \frac{s^2 - 9/4}{s^2} \left[
\frac{1}{[Q(s)]^{1/2}} - \frac{1}{[\frac{9}{2}(s^2-\bar\delta^2)]^{1/2}}
\right] ds.
\eeq
Thus, using (3.12)
\beq
|\Delta J| \leq \mbox{const} \int_{\bar\delta}^1 \frac{s}{s^2} \;
\frac{ds}{\sqrt{s^2 - \bar\delta^2}} = \frac{\mbox{const}}{\bar\delta}.
\eeq
Putting (3.31,32,33) together, there results
\beq
\frac{d \bar\tau}{d\bar\delta} = - \frac{3\sqrt{6}}{4} \; \frac{1}{\bar\delta}
+ O(1) \mbox{ as } \bar\delta \ra 0.
\eeq
Integrating (3.35), we obtain
\beq
\bar\tau(\bar\delta) = - \frac{3\sqrt{6}}{4} \ln \bar\delta + A' +
O(\bar\delta),
\eeq
for some constant $A'$.
Comparing with (3.29) we infer $A = A'$ and the proof of the estimate (3.5)
is complete. \QED

From $A = A'$ and (3.36) it is elementary to infer that
\beq
\bar\delta = \exp \left( - \frac{4}{3 \sqrt{6}} (\bar\tau - A)\right) +
O\left[\exp \left(- \frac{8}{3\sqrt{6}} \bar\tau \right) \right]
\mbox{ as } \bar\tau \ra \infty.
\eeq
Using (3.30), (3.35) can be written as
\beq
\frac{d \bar\tau}{d \bar C} = \frac{3}{4 \sqrt{2}} \; \frac{1}{\bar\delta^2}
+ O\left(\frac{1}{\bar\delta}\right).
\eeq

We want to evaluate the lapse $\alpha$ of the time function
$\tau = m \bar\tau$ along the central throat $r = r_C$. This, using
(2.23) and (2.24), is given by
\beq
\alpha(\tau) = \frac{1}{2m \bar\delta} \left(\frac{d \bar\tau}{d \bar C}
\right)^{-1}.
\eeq
Using (3.36) this finally leads to
\beqa
\alpha(\tau) &=& \frac{4}{3\sqrt{2}}\; \bar\delta +
O(\bar\delta^2) \no \\
&=& \frac{4}{3\sqrt{2}} \exp \left(\frac{4A}{3\sqrt{6}} \right)
\exp \left(-\frac{4\tau}{3 \sqrt{6}\;m} \right)
 + O \left[ \exp \left(- \frac{8\tau}{3 \sqrt{6}\;m}\right)\right] {}
     \nonumber\\
&&\mbox{ for } \tau \ra \infty.
\eeqa

We sum up our results in the 
\paragraph{Theorem:} For the chosen maximal foliation, with the
time function $\tau$  coinciding with proper time at infinity and
being zero on the time symmetric leaf $S$, the lapse along the central
geodesics orthogonal to the leaves behaves, as a function of $\tau$,
according to (3.40) with $A$ given by
\beq
A = \frac{3\sqrt{6}}{4} \ln |18(3\sqrt{2} - 4)| - 2 \ln \left|
\frac{3 \sqrt{3} - 5}{9 \sqrt{6} - 22}\right| = - 0.2181.
\eeq
It would be interesting to estimate the lapse for large $\tau$ along
arbitrary Eulerian observers rather than just the ones along $\Gamma$.
In terms of the coordinate $\rho$ introduced in \S~2 we conjecture that
\beq
\alpha(\rho,\tau) = B(\rho) \exp \left( - \frac{4}{3\sqrt{6}}\; 
\frac{\tau}{m}\right) + O\left[B^2(\rho) \exp \left( - \frac{8}{3\sqrt{6}}\;
\frac{\tau}{m} \right)\right],
\eeq
where $B(\rho)$ behaves for large $\rho$ as
\beq
B(\rho) \sim \mbox{const} \cdot \cosh \frac{4}{3 \sqrt{6}}
\left( \frac{\rho}{m}\right)^3.
\eeq
The form of $B(\rho)$ in Equ. (3.43) is motivated by the solution to the
lapse equation (2.25) on the limiting slice at $r = 3m/2$, which is
symmetrical with respect to the throat.

\appendix
\newcounter{zahler}
\renewcommand{\thesection}{\Alph{zahler}}
\renewcommand{\theequation}{\Alph{zahler}.\arabic{equation}}
\setcounter{zahler}{1}
\setcounter{equation}{0}
\section*{Appendix A}
The following discussion is similar in spirit to \cite{Gu}.
Let $\Sigma$ be the manifold ${\bf R} \times S^2$ with a Riemannian,
spherically symmetric metric, which we write in the ``radial'' gauge,
i.e.
\beq
g = d\ell^2 + r^2(\ell)d\Omega^2, \qquad r \in (0,\infty).
\eeq
The unit vector $\ell^i = (\partial/\partial \ell)^i$ is geodesic and
satisfies $(r' = dr/d\ell)$
\beq
D_i \ell_j = \frac{r'}{r} \; q_{ij},
\eeq
where $q_{ij} = g_{ij} - \ell_i \ell_j$ and prime means derivative w.r.t.
$\ell$. After a calculation, which most easily follows the lines of
Besse \cite{Bes}, we find for the Riemann tensor
\beq
R_{ijk\ell} \ell^j \ell^k = \frac{r''}{r} \; q_{i\ell}
\eeq
and
\beq
q_i^{i'} q_j^{j'} q_k^{k'} q_\ell^{\ell'} R_{i'j'k'\ell'} =
\frac{2}{r^2} (1 - r'{}^2) q_{k[i} q_{j]\ell}.
\eeq
Identities (A.3,4) imply that
\beqa
R_{ij} &=& - \frac{r''}{r} (2\ell_i \ell_j + q_{ij}) + 
\frac{1 - r'{}^2}{r^2} \; q_{ij} \\
R &=& - 4 \; \frac{r''}{r} + 2 \; \frac{1 - r'{}^2}{r^2}.
\eeqa
The extrinsic curvature on $\Sigma$, in order to be spherically symmetric,
has to be of the form
\beq
K_{ij} = v \;\ell_i \ell_j + w \;q_{ij}.
\eeq
The condition $K_{ij} g^{ij} = 0$ implies that $v + 2w = 0$. Using (A.2)
we have
\beq
D^i K_{ij} = \left(v' + 3 \; \frac{r'}{r} \; v \right) \ell_j.
\eeq
Thus the maximal momentum constraint implies $v = 2C/r^3$ for some
constant $C$. Consequently,
\beqa
K_{ij} &=& \frac{2C}{r^3} \; \ell_i \ell_j -  \frac{C}{r^3} \; q_{ij} \\
K_{ij} K^{ij} &=& 6 \; \frac{C^2}{r^6}.
\eeqa
Inserting (A.10) and (A.6) into the Hamiltonian constraint, there results
\beq
- 4 \; \frac{r''}{r} + 2 \; \frac{1 - r'{}^2}{r^2} = 6 \; \frac{C^2}{r^6}.
\eeq
Next we define $m(r)$ by
\beq
m(r) := \frac{r}{2} (1-r'{}^2) + \frac{C^2}{2r^3}.
\eeq
Now (A.11) implies that $dm/dr$ is zero. Thus
\beq
r' = \left( 1 - \frac{2m}{r} + \frac{C^2}{r^4}\right)^{1/2}.
\eeq
Assuming $m > 0$ and $0 \leq |C| < 3 \; \frac{\sqrt{3}}{4}\; m^2$, 
there are two initial-data sets
consistent with (A.9) and (A.13). One starts at $r = 0$, expands to an
$r_{\rm max} < 3m/2$ and collapses back to $r = 0$. The other is an
asymptotically flat complete metric on ${\bf R} \times S^2$ with mass $m$
at both ends which is symmetric with respect to the throat at
$r = r_C > 3m/2$ with
\beq
1 - \frac{2m}{r_C} + \frac{C^2}{r^4_C} = 0.
\eeq
Here we restrict ourselves to asymptotically flat data. These constitute
a 2-parameter family of solutions to the spherically
symmetric, maximal vacuum constraints. Of course, we know from the
Birkhoff theorem that members of this family with different $C$ but the
same $m$ have all to lie in the same spacetime, namely the extended
Schwarzschild spacetime. ``Discovering'' this fact in the present context
amounts to finding the ``height function'' written down in Sect.~2. The
trick is to try to find the remaining Killing vector and to seek the
$\Sigma_C$'s as graphs over the surfaces orthogonal to this Killing
vector. If $(g_{ij},K_{ij})$ evolve to a spacetime having another Killing
vector, there must be a function $N$, not identically zero, and a vector
field $X^i$ so that
\beq
2N K_{ij} + 2 D_{(i} X_{j)} = 0.
\eeq
Assuming $X^i$ to be again spherical, i.e.
\beq
X_i = \mu \; \ell_i, \qquad \mu = \mu(r),
\eeq
and again using (A.2) and (A.9) we infer that
\beqa
- 2 \; \frac{NC}{r^3} + 2r'\; \frac{\mu}{r} &=& 0 \\
4 \; \frac{NC}{r^3} + 2r' \; \frac{d\mu}{dr} &=& 0.
\eeqa
After combining (A.17) and (A.18), there results
\beqa
\mu(r) &=& \frac{D}{r^2}, \qquad D = \mbox{const} \\
N &=& \frac{D}{C} \; r',
\eeqa
where we have assumed $C \neq 0$. We assume without loss that $D = C$. 
The existence of $(N,X^i)$ solving Equ.'s (A.15) does not necessarily imply
that the vacuum evolution of the initial data set has a static Killing
vector. There also has to be satisfied
\beq
\cL_X K_{ij} + D_i D_j N = N(R_{ij} - 2K_{i\ell} K_j^\ell).
\eeq
It is straightforward to check that (A.19,20) {\bf do} satisfy (A.21).
(In the case where $C$ is zero, $X^i = 0$ and Eq.(A.21) implies that
$N \sim r'$.)

We remark in passing that the function $N$, by virtue of (A.15) and 
(A.21), satisfies
\beq
\Delta N = N K_{ij}K^{ij}.
\eeq
(Of the two linearly independent spherical solutions of (A.22) $N$
is that combination which vanishes on the throat.)

It now follows that for $r > r_C$ the metrics
\beq
ds^2 = - (N^2 - g_{ij}X^i X^j)d\sigma^2 + 2g_{ij} X^j dx^i d\sigma +
g_{ij}dx^i dx^j,
\eeq
with $N_i$, $X^i$, $g_{ij}$ extended in a 
$\sigma$-independent way to ${\bf R} \times \Sigma$, are vacuum solutions
evolving from the above initial data sets. They have $\xi^\mu =
(\partial/\partial \sigma)^\mu$ as a Killing vector. More explicitly,
since
\beq
N^2 - X_i X^i = 1 - \frac{2m}{r},
\eeq
we have
\beq
ds^2 = - \left(1 - \frac{2m}{r}\right) d\sigma^2 + 2 \; \frac{C}{r^2} \;
d\ell d\sigma
+ d\ell^2 + r^2 d\Omega^2,
\eeq
where $r(\ell)$ is given implicitly by
\beq
\ell(r) = \int_{r_C}^r \frac{dx}{\sqrt{1 - 2m/x + C^2/x^4}}.
\eeq
(For $C = 0$, $\ell(r)$ can be written as
$\ell(r) = r \sqrt{1 - 2m/r} + m \ln \left| \frac{1 + \sqrt{1 - 2m/r}}
{1 - \sqrt{1 - 2m/r}}\right|$.)

Note that for $C \neq 0$ the above metrics extend smoothly across $r = 2m$.
We now seek a function $t$ with level surfaces orthogonal to
$\partial/\partial \sigma$. Writing this function as
\beq
t = F(r) + \sigma,
\eeq
we obtain from
\beq
g_{\mu\nu} \xi^\mu dx^\nu = - (N^2 - X_i X^i)d\sigma + X_i dx^i =
\omega(dF + d\sigma),
\eeq
for some function $\omega$, the equation
\beq
- D_i F = \frac{X_i}{N^2 - X_j X^j}
\eeq
which makes only sense off the horizon. Using (A.16,19) this leads to
\beq
\frac{dF}{dr} = - \frac{C}{r^2 - 2mr} \frac{1}{\sqrt{1 - 2m/r + C^2/r^4}}.
\eeq
Now consider the coordinate transformation
\beq
\sigma = t - F.
\eeq
Then
\beq
ds^2 = - (N^2 - X_j X^j)dt^2 + \bar g_{ij} dx^i dx^j,
\eeq
with
\beqa
\bar g_{ij} &=& g_{ij} + 2X_{(i} F_{,j)} - (N^2 - X_\ell X^\ell)
F_{,i} F_{,j} \\
&=& g_{ij} + (N^2 - X_\ell X^\ell)^{-1} X_i X_j ,
\eeqa
where $X_i := g_{ij} X^j$. Using (A.29) and (A.30),
\beq
\bar g_{ij}dx^i dx^j = \left[ 1 + \left( 1 - \frac{2m}{r}\right)^{-1}
\frac{C^2}{r^4}\right] d\ell^2 + r^2 d\Omega^2 =
\left( 1 - \frac{2m}{r}\right)^{-1} dr^2 + r^2 d\Omega^2.
\eeq
We have thus recovered the Schwarzschild metric. In particular this 
calculation shows that the parameter $C$ in our initial-data sets is
``pure gauge'': initial data with the same $m$ lie in the same spacetime,
namely as level sets of the function $\sigma$. They can also be written as
\beq
t = F_C(r)
\eeq
and its translates under $\xi^\mu = (\partial/\partial t)^\mu$, where
\beq
F_C(r) = - C \int_{r_0}^r \frac{dx}{(1 - m/x)(1 - 2m/x + 
C^2/x{}^4)^{1/2}}
\eeq
for some $r_0$. Taking $r_0 = r_C$ we have, with $h(r,C) = F_C(r)$,
recovered (2.12).

It is shown in Sect. 2 that
\beq
t = h(r,C)
\eeq
implicitly defines a smooth time function on the $r > 3m/2$-subset of
the future half of Kruskal. The boost function $N$ obtained in this
Appendix satisfies the same equation, on each leaf
$\Sigma_C$ as the lapse function of $C$, namely (A.22). 
The reason for this is that,
for fixed $\Sigma_C$, $\xi^\mu$ defines another local foliation, which
is again maximal since $\xi^\mu$ is a Killing vector.

\addtocounter{zahler}{1}
\setcounter{equation}{0}
\section*{Appendix B}
Consider
\beq
F(x,E) = \int_{x_E = V^{-1}(E)}^x \frac{W(y)}{[E - V(y)]^{1/2}} dy,
\eeq
where $V$ is a smooth function $V : [x_0,\infty) \ra {\bf R}$ with
\beq
0 < V(x_0), \qquad V'(x) < 0 \mbox{ for } x > x_0, \qquad V(\bar x) = 0
\eeq
and
\beq
0 < E < V(x_0).
\eeq
The function $W$ is smooth except perhaps at $x = \bar x$, where it may 
have a simple pole. (Thus the pole of $\sqrt{E}\;W(y)/[E-V(y)]^{1/2}$ 
is independent of $E$.)
In the latter case (B.1) is to be understood in the principal-value sense 
and the following operations valid for $x \neq \bar x$. Next define
(we follow \cite{Ch} in spirit)
\beq
J(x,E) = \int_{x_E}^x [E - V(y)]^{1/2} V(y) W(y) dy.
\eeq
Note that $V \cdot W$ is smooth. Equ. (B.4) can be rewritten as follows
\beq
J(x,E) = - \frac{2}{3} \int_{x_E}^x \frac{d}{dy} [E - V(y)]^{3/2}
\frac{V(y)W(y)}{V'(y)} dy, 
\eeq
\beq
J(x,E) = - \frac{2}{3} [E - V(x)]^{3/2} \frac{V(x) W(x)}{V'(x)} 
 + \frac{2}{3} \int_{x_E}^x [E - V(y)]^{3/2} \frac{d}{dy}
\left[ \frac{V(y) W(y)}{V'(y)} \right] dy.
\eeq
Differentiating (B.6) w.r.t. $E$ twice, we obtain
\beq
\frac{d^2}{dE^2}\; J(x,E) = - \frac{1}{2} \frac{1}{[E - V(x)]^{1/2}}
\frac{V(x) W(x)}{V'(x)} 
+ \frac{1}{2} \int_{x_E}^x \frac{1}{[E - V(y)]^{1/2}} \frac{d}{dy}
\left[ \frac{V(y) W(y)}{V'(y)} \right] dy.
\eeq
On the other hand, differentiating (B.4) once w.r.t. $E$ it follows that
\beqa
\frac{d}{dE} J(x,E) &=& \frac{1}{2} \int_{x_E}^x 
\frac{V(y)}{[E - V(y)]^{1/2}} W(y) dy \no \\
&=& \frac{1}{2} \int_{x_E}^x \frac{V(y) - E + E}{[E - V(y)]^{1/2}}
W(y) dy \no \\
&=& - \frac{1}{2} \int_{x_E}^x [E - V(y)]^{1/2} W(y) dy +
\frac{1}{2} E F(x,E).
\eeqa
Differentiating (B.8) once more w.r.t. $E$ and comparing with (B.7) we
finally find
\beqa
\lefteqn{\frac{1}{4} F(x,E) + \frac{1}{2} E \frac{d}{dE} F(x,E) = } \no \\
&=& - \frac{1}{2} \frac{1}{[E - V(x)]^{1/2}} \frac{V(x) W(x)}{V'(x)}
+ \frac{1}{2} \int_{x_E}^x \frac{1}{[E - V(y)]^{1/2}} \frac{d}{dy}
\left[ \frac{V(y) W(y)}{V'(y)} \right] dy.
\eeqa
In our case we will have that $V'(x_0) = 0$ and we study the blow-up of
$F_\infty(E) = \lim_{x \ra \infty} F(x,E)$ as $E$ tends to
$E_0 = V(x_0)$. As for a mechanical analogue, we could think of a particle
on a half-line in a repulsive potential $V(x)$ and imagine $F(x,E)$
to be the time it takes a particle of energy $E$ to travel from $x_0$
to $x$. (If it were not for the presence of $W(y)$ in (B.1), this 
interpretation would be literally true.)
The force on the particle grows so fast for large $x$, that the
particle reaches infinity in finite time $F_\infty(E)$. There is an
unstable equilibrium point at $x = x_0$. We ask for the way in which
$F_\infty(E)$ blows up as $E$ approaches $V(x_0)$. If the energy $E$ is
further increased,the orbits reach $x = 0$: this corresponds to maximal
slices hitting the singularity.

To make contact with our function $h(r,C)$, set
$$
V(x) = - x^4 + 2mx^3, \qquad E = C^2, \qquad 
W(x) = - \frac{1}{1 - 2m/x},
$$
\beq
h(r,C) = C F(r,C^2),
\eeq
$$
x_0 = \frac{3m}{2}, \qquad \bar x = 2m.
$$
Thus
\beq
\frac{d}{dC} h(r,C) = 2E \left. \frac{d}{dE} F(r,E)\right|_{E=C^2}
+ F(r,C^2)
\eeq
which, combined with (B.9), gives
\beq
\frac{d}{dC} h(r,C) = \frac{1}{2 (r - 3m/2)  \sqrt{1 - 2m/r + C^2/r^4}}
- \frac{1}{2} \int_{r_C}^r \frac{x(x-3m)dx}
{(x - 3m/2)^2(x^4 - 2mx^3 + C^2)^{1/2}}.
\eeq

\section*{Acknowledgements}
The authors thank Piotr Chru\'sciel for helpful discussions. One of us
(N. \'O M.) gratefully acknowledges the support and hospitality of ESI.

\end{document}